\documentclass[aip,twocolumn,letterpaper]{revtex4}

\usepackage{fullpage}

\usepackage{graphics}
\usepackage{epsfig}

\begin{document}
\title{The tritium burn-fraction in DT fusion}
\author{Allen H. Boozer}
\affiliation{Columbia University, New York, NY  10027\\ ahb17@columbia.edu}

\begin{abstract}

The fraction of the tritium that is burned during one pass through a DT fusion system, $f_{tb}$, is a central issue for success of fusion energy.  Reducing the tritium fraction, $f_t$, in a $DT$ burning plasma below a half increases the burn fraction, $f_{tb}\propto1/f_t$ but also the required confinement to achieve a burn $nT\tau_E\propto1/f_t(1-f_t)$.  A doubling of the fractional burn entails only a 4/3 enhancement of the required $nT\tau_E$.  The energy confinement time $\tau_E$ in tokamaks and stellarators is empirically gyro-Bohm with an approximate factor of two between the best and worse results used to construct scaling laws.  Gyro-Bohm is also the approximate level of transport needed in a power plant.  What has received little study are $\tau_t/\tau_E$, the ratio of the tritium to the energy confinement time, and $\tau_\alpha/\tau_E$, the ratio of the alpha particle to the energy confinement time.  The tritium burn fraction is proportional to  $\tau_t/\tau_E$, so the larger the better.  The contamination of the plasma by helium ash is proportional to $\tau_\alpha/\tau_E$, so the smaller the better.

\end{abstract}

\date{\today} 
\maketitle

The fraction of the tritium that is burned during one pass through a DT fusion system is a central issue for success of fusion energy \cite{Baylor:2017,Tritium:2021}.  This note gives a simple calculation of the burn-up fraction, simplified even from \cite{GA burn-frac}, and of the alpha ash accumulation.

The burn-time of the tritium $\tau_{tb}$ is defined by
 \begin{eqnarray} 
 \frac{dn_t}{dt} &=& - n_tn_d\langle\sigma v\rangle_{dt}, \mbox{   so} \label{dn_t/dt} \\
 n_t &\propto& e^{-t/\tau_{tb}} \mbox{    with  } \\
 \tau_{tb} &\equiv& \frac{1}{n_d\langle\sigma v\rangle_{dt}}.
 \end{eqnarray}  
 The quantity $\langle\sigma v\rangle_{dt}$, which has units of meters-cubed divided by time, gives the probability that Maxwellian deuterium and tritium ions will fuse as a function of their temperature.
 
 The fraction of the tritium burned before it is swept out of the plasma on the tritium confinement time $\tau_t$ is
 \begin{equation}
 f_{tb} \equiv 1- e^{-\tau_t/\tau_{tb}} \approx \frac{\tau_t}{\tau_{tb}}.
 \end{equation}
 since the tritium-burn fraction $f_{tb}$ is much less than unity.  The fractional burn relation, $f_{tb}=\tau_t/\tau_{tb}$, will be treated as if it is exact.
 
 The power density of the plasma heating by the alpha particles $p_\alpha$ and the energy confinement time $\tau_E$ required to maintain a burning plasma are
 \begin{eqnarray}
 p_\alpha &=& \mathcal{E}_{\alpha} n_tn_d\langle\sigma v\rangle_{dt}; \\
 \tau_E &\equiv& \frac{3 n k_BT}{ \mathcal{E}_{\alpha} n_tn_d\langle\sigma v\rangle_{dt}}, \label{tau_E}
 \end{eqnarray}
 where $\mathcal{E}_{\alpha}\approx3.5~$Mev is the energy release in an alpha particle per reaction, $n=n_d+n_t$ and $T$ is the plasma temperature.  The factor $3nk_BT$ is the thermal energy of the electrons plus the ions, where $k_B$ is the Boltzmann constant. 
 
 The ratio of the required energy confinement time to the burn-time of the tritium is then
 \begin{eqnarray}
\frac{ \tau_E}{\tau_{tb}} = \frac{3  k_BT }{ \mathcal{E}_{\alpha} f_t}, \mbox{   where   }
f_t &\equiv& \frac{n_t}{n}
\end{eqnarray}
is the fraction of the ions that are tritium.  The fraction of the tritium burned is then
\begin{equation}
f_{tb} = \frac{3 k_B T }{ \mathcal{E}_{\alpha}}\frac{1}{f_t} \frac{\tau_t}{\tau_E}.
\end{equation}
The smallness of the tritium-burn fraction is primarily due to $3k_BT/\mathcal{E}_{\alpha}$ being of order one percent.

The tritium-burn fraction can be made arbitrarily large by making the tritium fraction small at the cost of making the required energy-confinement time longer;
\begin{eqnarray}
 n\tau T &\equiv& \frac{3}{f_t(1-f_t)} \frac{k_BT^2}{\mathcal{E}_{\alpha}\langle\sigma v\rangle_{dt}}  \\
 &=& \frac{ \Big( n\tau T \Big)_{min}}{4f_t(1-f_t)},
 \end{eqnarray}
 where $( n\tau T )_{min}$ is the minimum-confinement requirement for achieving a burn.  Halving the tritium fraction and doubling its burn fraction requires only a factor of 4/3 enhancement of $(n\tau T )_{min}$; quadrupling the burn fraction requires a factor of $16/7\approx2.29$ enhancement.
 
The helium ash from the alpha particles will degrade the fusion properties of the plasma unless the confinement time of alpha particles $\tau_\alpha$ is only moderately longer than the energy confinement time.  Since an alpha particle is created for each tritium that is burned, $dn_\alpha/dt=-dn_t/dt$. Equations (\ref{dn_t/dt}) and (\ref{tau_E}) imply
\begin{eqnarray}
\frac{n_\alpha}{n}&=&\frac{n_tn_d}{n}\langle\sigma v\rangle_{dt}\tau_\alpha \\
&=&\frac{3k_BT}{\mathcal{E}_{\alpha}} \frac{\tau_\alpha}{\tau_E}.
\end{eqnarray}

The contribution of alpha particles to the energy density is $(9/2)n_\alpha k_BT$ and the contribution to the pressure is $3n_\alpha k_BT$  since each alpha adds  three particles: one ion and two electrons. 
Bremsstrahlung is proportional to $Z_{eff}\equiv \sum n_ZZ^2/n_e$, so the enhancement  of the helium ash of the bremsstrahlung power loss is four times higher that its contribution to the number density of ions $n_\alpha$.

It is important that the fractional burn $f_{tb}$ be large and the alpha dilution $n_\alpha/n$ be small.  The relation 
 \begin{equation}
 f_{tb}=\frac{\tau_t}{\tau_\alpha}\frac{n_\alpha/n}{f_t}
 \end{equation} 
 implies a reduced tritium faction gives a larger fractional burn at the maximum tolerable dilution.

The energy confinement time $\tau_E$ in tokamaks and stellarators is empirically gyro-Bohm \cite{CO2-Stell} with an approximate factor of two between the best and worse results used to construct scaling laws; see Figure 4 in \cite{W7-X:Nature}.  

 Gyro-Bohm is also the approximate level of energy transport needed in a power plant \cite{CO2-Stell}.  However, the Greenwald density limit requires tokamaks to operate at a much higher temperature \cite{Zohm:pp} than stellarator power plants to achieve an adequate power density, which because of the temperature scaling of $\langle\sigma v\rangle_{dt}$ and of gyro-Bohm transport requires a large reduction in the transport relative to gyro-Bohm \cite{Boozer:fast-path}.  
 
 An energy confinement time  $\tau_E$ that is too short can be compensated by increasing the magnetic field and plasma size.  However, a $\tau_t/\tau_E$, the ratio of the tritium to the energy confinement time, that is too short or a $\tau_\alpha/\tau_E$, the ratio of the alpha particle to the energy confinement time, that is too long are probably more difficult to modify.  It is important that the physics that determines these ratios be studied both theoretically and empirically.


\section*{Acknowledgements}

This material is based upon work supported by the U.S. Department of Energy, Office of Science, Office of Fusion Energy Sciences under Award Number DE-FG02-03ER54696.


\end{document}